\documentstyle[aps,multicol,graphicx]{revtex}
\begin{document}
\def\R{\textrm{I}\!\textrm{R}}
\def\G{{\cal G}}
\title{Generalized entropies from first principles}

\author{M. P. Almeida}

\address{Departamento de F\'{\i}sica, Universidade Federal do Cear\'a,
60455-760 Fortaleza, Cear\'a, Brazil}
\maketitle

\begin{abstract}
It is presented a derivation of power law canonical distributions from
first principle statistical mechanics, including the exponential 
distribution as a particular case. It is  shown that these distributions 
arise naturally, and that the heat capacity of the heat bath is the 
condition that determines its type.  As a consequence, it is given  a physical 
interpretation for the parameter $q$ of the generalized entropy.      
\end{abstract}

%\begin{multicols}{2}

In a 1988 paper Tsallis\cite{tsallis88} proposed a generalized entropy 
given by the form
\begin{equation}\label{Sq}
S_q=k{1 \over q-1}\left[1-\sum_{i=1}^W p_i^q\right],
\end{equation}
with $k$ a positive constant, $q$ a parameter and $\{p_i,i=1,\ldots,W\}$
a discrete probability distribution over the states labeled by $i$. 
It is easily observed that $S_q$ in the limit $q\to 1$
reduces to the standard Boltzmann-Gibbs (BG) entropy
\begin{equation}\label{S}
S=-k\sum_{i=1}^Wp_i\ln p_i.
\end{equation}
Based on this entropy Eq.(\ref{Sq}), a wealthy of papers 
(see Tsallis\cite{www} for an updated list) has been presented 
developing 
an alternative thermodynamical formalism and applying it to actual physical
systems. 
 
The supporters of the generalized  thermodynamics 
argue that the BG approach must have a restricted domain of applicability, and
that it satisfactorily describes short-range (in time and space) systems
with non (multi)fractal boundary conditions \cite{tsallis99}, and to reinforce
this argument they quote a list of physical systems that present a behavior 
best described by the generalized thermodynamics (see \cite{tsallis99} and the 
references therein). 
There are recent convincing  examples in favor of the generalized 
thermodynamics in a variety of fields such as turbulence \cite{Arimitsu,Beck}, 
elementary particles \cite{Bediaga}, and anomalous diffusion of micro-organisms
in cellular aggregates \cite{Upadhyaya}.
However, more experimental and theoretical work is still relevant to assure the 
applicability and robustness of this generalized approach.

In the large majority of papers that advocate the generalized 
thermodynamics, the starting point for the analysis is the adoption of the 
$S_q$ entropy (Eq.(\ref{Sq})). Up to now, little has been proposed in 
terms of {\it first principles} to explain the appearance of this entropy
form in physical systems. But first principles techniques have already been 
deployed by S. Abe and A. K. Rajagopal \cite{Abe1} to derive and study power law 
canonical distributions from microcanonical distributions. 
Their development results from the use
of generalizing functions for the logarithm and for the exponential in classical 
statistical mechanics procedures 
\cite{Kubo}.   

 There are several works \cite{Lyra} where the parameter $q$ is linked to
the system sensibility on the initial conditions. But a complete specification  
of its physical meaning is still an open question. 

In this letter, we follow the procedure of derivation of the 
canonical  distribution, as presented in standard statistical mechanics 
textbooks, viz.,
\cite{Kubo,Kinchin,Weiner}, to arrive at an understanding of the similarities 
and 
dissimilarities between the BG (exponential) canonical distribution and its 
generalized counterpart, which is a power law. The arguments used 
are exact and involve no approximation mechanism. From it, we can see that the 
power law and the exponential distributions are members of a family of 
possible canonical
distributions that arises naturally from statistical mechanics. The signature
of these distributions being determined by a parameter that has macroscopic
identification.  
From this simple and basic development, we can get a clear
understanding of the physical meaning of the parameter $q$.
  
In standard statistical mechanics textbook the derivation of the canonical 
distribution for a system $\G_1$ 
interacting with heat bath $\G_2$ is based on the structure functions, 
$\Omega_2$ of $\G_2$ and $\Omega$ of the whole system $\G=\G_1+\G_2$, 
which expresses the number of configurations of the phase space contained
in a shell of constant energy, and is given by
\begin{equation}
{\Omega_2(a-E_1)\over \Omega(a)}={\Omega_2(a)\over\Omega(a)}
{\Omega_2(a-E_1)\over \Omega_2(a)}.
\end{equation}
Taking the Taylor expansion of $\Omega_2(a-E_1)$ at $E=a$ and defining 
$\beta=\Omega_2'/\Omega_2$, it can be easily verified that the specification
of a constant value for the derivative of the inverse of $\beta$ with respect 
to the energy, viz., 
\begin{equation}\label{dbeta}
{d\over dE}\left({1\over \beta}\right)=q-1,
\end{equation}
is sufficient to determine the
form of the function $\Omega_2(a-E_1)/ \Omega_2(a)$.
Indeed, when 
$d(1/\beta)/dE=0$ we get $\Omega_2(a-E_1)/ \Omega_2(a)=\exp(-\beta E_1)$, and
when $d(1/\beta)/dE=q-1\ne 0$ we get   $\Omega_2(a-E_1)/ \Omega_2(a)=
[1-(q-1)\beta E_1]^{1/(q-1)}$, for $ 1-(q-1)\beta E_1>0$, and zero otherwise, 
which is a power law that we denote by $e_q(-\beta E_1)$.
This last function may be considered as a generalized
exponential, in the sense that in the limit $q\to 1$, $e_q(x) \to \exp(x)$.
Also, in classical statistics mechanics 
$1/\beta$ is proportional to the 
temperature $(1/\beta=kT)$.  
Therefore the parameter $q$ is related
to the energy derivative of the heat bath temperature.
It measures the temperature sensitivity of the heat bath to energy variations.
Also, we see that for the case $q=1$, which results in the (BG) exponential
distribution, 
\begin{equation}
{d\over dE}\left({1\over \beta}\right)=0
\end{equation}
which means that the BG canonical distribution has the implicit condition
that temperature of the heat bath does not depend on its energy.
This is realized in the case of an infinity heat bath, which can gain or lose
any amount of energy without changing the temperature, which is equivalent
to saying that it has infinite heat capacity. 
For the case of finite heat bath, this condition should no longer be valid
and we should get power law canonical distributions.

In terms of the structure function $\Omega_2$ this condition implies that
$\Omega_2(E)= C_1(E+C_2)^{1/q-1}$ when $q\neq 1$, and 
$\Omega_2(E)=C\exp(\alpha E)$ when $q=0$.
It is worthing noting that in some textbooks, e.g. 
Uhlenbeck and Ford\cite{uhlenbeck},
the power law distribution appears in an intermediate step  in the derivation 
of the exponential canonical distribution, which is obtained by letting the
system size go to infinity.

Observe also that the derivative of the temperature $(1/\beta)$ 
with respect to energy is proportional to the inverse of the heat capacity, 
which leads to the 
following analysis: 
For $q>1$, the derivative in Eq.(\ref{dbeta}) becomes positive and 
this  situation represents a heat bath with a finite positive heat capacity. 
Analogously, when $q<1$ the derivative of temperature with energy becomes 
negative and the heat bath has a negative finite heat capacity.

Let's make a comparison between statistical mechanics and continuum 
thermodynamics entities
in order to derive a microscopic expression for the entropy. 
The details of the following can be found in \cite{Weiner}.
Let $\rho=\rho({\bf q},{\bf p})$ be the canonical distribution over the phase space 
of a 
given system. Let's consider the case where the macroscopic internal energy 
has the Hamiltonian as its microscopic counterpart and is given by
\begin{equation}
U=\int_\Gamma H\rho d{\bf q} d{\bf p}.
\end{equation}
In a quasi-static process we can compute the rate of change of $U$ with 
respect to time in the form
\begin{equation}
\dot{U}=\int_\Gamma \dot{H}\rho d{\bf q} d{\bf p} + \int_\Gamma H\dot{\rho} d{\bf q}
d{\bf p},
\end{equation}
where the dot on top of the functions represents the total derivative
with respect to time.
This expression represents the first law of thermodynamics: 
$\dot{U}=\dot{W}+\dot{Q}$,
with $\dot{W}=\int_\Gamma \dot{H}\rho d{\bf q} d{\bf p}$ and $\dot{Q}=\int_\Gamma H
\dot{\rho} d{\bf q} d{\bf p}$.
The second law of thermodynamics asserts that there are
functions $T(\beta)$, absolute temperature scales, and $S$, the entropy, such 
that  
$dQ=T(\beta)dS$.
In the following paragraph we have a prescription for the determination
of entropy functions from the canonical distribution. This establishes
a one-to-one correspondence between canonical distributions and entropy forms,
in the sense that an entropy form determines a canonical distribution via
a {\it constrained maximum entropy} problem, and the canonical distribution
determines (up to constant multiplication and addition) the entropy form.

Under the hypothesis that the canonical distribution $\rho=f(H)$ is 
a monotonic function of $H$, such that we can find $H$ as the inverse 
function $H=f^{-1}(\rho)$, we can define a function
\begin{equation}\label{F}
F(\rho)={1\over T(\beta)} \int_0^\rho f^{-1}(\xi)d\xi +\psi\rho,
\end{equation}
with an arbitrary constant $\psi$, such that 
\begin{equation}\label{ent1}
S=\int_\Gamma F(\rho) d{\bf q} d{\bf p}
\end{equation}
is an entropy function, i.e., we have $\dot{Q}=T(\beta)\dot{S}$.
Indeed, from these definitions we have that
\begin{eqnarray*}
\dot{S}&=&\int_\Gamma F'(\rho)\dot{\rho} d{\bf q} d{\bf p}\\
&=&{1\over T(\beta)} \int_\Gamma H\dot{\rho} d{\bf q} d{\bf p} +
\psi\int_\Gamma  \dot{\rho} d{\bf q} d{\bf p} \\
&=& {\dot{Q}\over T(\beta)},
\end{eqnarray*}
where we used the fact that $\int_\Gamma\dot{\rho} d{\bf q} d{\bf p} =0$. 

In the case we have 
$
\rho({\bf q},{\bf p})=\exp(-\beta H({\bf q},{\bf p}))=f(H),
$
we get that 
\begin{equation}
F(\rho)= -{1\over \beta T(\beta)}[\rho\ln(\rho) -\rho] +\psi\rho,
\end{equation}
which, by taking $T(\beta)=1/\beta$ and $\psi = -1$, 
produces the Boltzmann-Gibbs entropy 
$S=-\int_\Gamma \rho\ln(\rho)d{\bf q} d{\bf p}$.

For $q\ne 1$ and 
$
\rho({\bf q},{\bf p})=e_q(-\beta H)
$
we have that
\begin{equation}
F(\rho)= {1\over q\beta T(\beta)}\left({\rho-\rho^q \over q-1}\right) +
\left[{1\over q\beta T(\beta)} +\psi\right] \rho,
\end{equation}
from which, taking $T(\beta)=1/(q\beta)$ and $\psi=-1$, we get the generalized
entropy 
\begin{equation}
S_q=\int_\Gamma{\rho-\rho^q\over q-1}d{\bf q} d{\bf p}
\end{equation}  

From the above, we can conclude that the existence of power law canonical 
distributions is perfectly justified from first principles, and that the 
exponential distribution may be considered as a particular case, the 
determining feature being the derivative $d(1/\beta)/dE$ of the heat bath 
temperature. This also implies that the generalized entropies are the adequate
entropy forms for the case $d(1/\beta)/dE\ne 0$, which may be considered
as a condition of heat bath with finite heat capacity. The Boltzmann-Gibbs 
entropy is the right form for systems with ideal heat bath.  

\vskip0.5cm
 
\noindent{I would like to thank Profs. J. E. C. Moreira, and R. N. Costa Filho for
their comments and suggestions. 
I want to thank also Prof. C. Tsallis for his comments and 
for bringing  my  attention to some recent publications about the topic. 
This work was partially supported by  CNPq (Brazil)}.

%\end{multicols}
\end{document}